\documentclass[prd,aps, preprintnumbers,showpacs,twocolumn, nofootinbib,superscriptaddress,notitlepage]{revtex4-1}
\usepackage[normalem]{ulem}
\usepackage{epsfig}
\usepackage{amsfonts}
\usepackage{amsmath}
\usepackage{slashed}
\usepackage{graphicx}
\usepackage{color}
\usepackage{mathtools}
\usepackage{subfigure}
\usepackage{graphicx}
\usepackage{hyperref}
\usepackage{stmaryrd}
\usepackage{amssymb,psfrag}
\usepackage{caption}
\allowdisplaybreaks[4]

\newcommand{\beq}{\begin{eqnarray}}
\newcommand{\eeq}{\end{eqnarray}}

\newcommand{\nn}{\nonumber}

\def\slash#1{#1 \hskip-0.45em /}

\begin{document}

\title{Logarithmic moments of $B$-meson quasidistribution amplitude}

\author{Shu-Man Hu}
\affiliation{School of Physics and Microelectronics, Zhengzhou University, Zhengzhou, Henan 450001, China}

\author{Ji Xu}
\email{xuji\_phy@zzu.edu.cn}
\affiliation{School of Physics and Microelectronics, Zhengzhou University, Zhengzhou, Henan 450001, China}

\author{Shuai Zhao}
\email{zhaos@tju.edu.cn}
\affiliation{Department of Physics, Tianjin University, Tianjin 300350, China}

\begin{abstract}
  It was demonstrated that the lattice simulation of $B$-meson light-cone distribution amplitude (LCDA) is feasible via the quasi-distribution amplitude (quasi-DA) in large momentum effective theory (LaMET).
  The structures of  logarithmic moments (LMs) of $B$-meson quasi-DA is explored in this work. The one-loop results indicate mixing in the matching: the $n$-th LM would be not only factorized into the $n$-th LM of LCDA, but also other moments with different power, accompanied by short distance coefficients. These results supply the understanding of the matching  in LaMET and may provide guidance to the lattice study of LMs or other parameters of $B$-meson LCDA.
\end{abstract}

\maketitle

\section{Introduction}
\label{Introduction}
The $B$-meson light-cone distribution amplitudes (LCDAs) serve as a fundamental quantity for characterizing the internal structure of $B$-mesons in terms of their constituent quarks and gluons. Initially introduced to capture the essence of generic exclusive $B$-decays, these distribution amplitudes have since played a pivotal role in the development of factorization theorems \cite{Szczepaniak:1990dt,Grozin:1996pq,Beneke:2000wa,Beneke:1999br,Beneke:2001ev,Li:2020rcg,Feldmann:2014ika,Hua:2020usv}.

In the realm of numerous hard exclusive reactions, the factorization theorem highlights the significance of the inverse moment (IM) of the LCDA, particularly in leading-twist contributions. Notably, the IM holds crucial phenomenological relevance, governing leading-power spectator interactions in diverse processes such as leptonic decays ($B\to \gamma \ell \nu$) \cite{Wang:2018wfj}, semileptonic decays ($B\to \pi \ell \nu$) \cite{Khodjamirian:2011ub}, and hadronic decays ($B\to\pi\pi$) \cite{Lee:2005gza}. Additionally, the IM plays a crucial role in constructing models for LCDA \cite{Korchemsky:1999qb,Wang:2015uea,Meissner:2013pba}.

When the analysis of $B$-meson decays extends beyond the tree level, the logarithmic moments (LMs) become essential, particularly in precision studies such as $B\to \gamma \ell \nu$, where they dominate theoretical errors \cite{Beneke:2011nf}. This emphasizes the critical role that both IM and LMs play in advancing our understanding of $B$-meson decays and underscores their importance in theoretical modeling and precision calculations.

Despite the crucial significance of  IMs  and LMs, our understanding of them remains limited. This is primarily due to their encoding of information on nonperturbative dynamics, making their computation challenging from the first principles of  QCD. Existing results on IM and LMs are largely model-dependent, lacking satisfactory constraints. This limitation hampers the precision of theoretical predictions in relevant studies within $B$ physics. Consequently, there is a clear imperative to prioritize the determination of these moments in a model-independent manner, addressing a critical gap in our knowledge and advancing the field of $B$ physics.
Nonperturbative methods such as lattice QCD offers an alternative way out, the continuum HQET community would like a simulation on IM and LMs on the lattice. However, a practical difficulty is that these moments are defined in terms of the bilocal operators with lightlike separation which cannot be related to local operators, making the direct lattice simulation on Euclidean space essentially unfeasible.

In the last decade, it was pointed out that this difficulty can be overcome by employing the large-momentum effective theory (LaMET)~\cite{Ji:2013dva,Ji:2014gla} (see also~\cite{Ji:2020ect,Zhao:2018fyu,Cichy:2018mum} for reviews ). This is realized by simulating appropriately chosen equal-time correlations on the lattice and then converting them to the desired physical quantities.  In addition to LaMET, other related proposals such as pseudodistributions~\cite{Radyushkin:2017cyf,Radyushkin:2019mye} and lattice cross sections ~\cite{Ma:2017pxb,Li:2020xml} also made many progresses in accessing parton physics on the lattice. In the past few years, the ideas of lattice parton physics have also been applied to the structure of heavy hadrons~\cite{Jia:2015pxx,Kawamura:2018gqz,Wang:2019msf,Zhao:2020bsx,Xu:2022guw,Hu:2023bba}.
The IM of $B$-meson quasi-DA has been introduced and studied in~\cite{Xu:2022krn}. It was found that the IM of quasi-DA can be factorized into IM and the first two LMs of LCDA. This implies the existence of a mixing matrix which connects moments of quasi-DA and moments of LCDA. Furthermore, the study of moments of the quasi-DA in itself is of interest, as it provides valuable insights into the characteristics of perturbative matching in LaMET. In this work, we  extend the investigation to include the LMs of the $B$-meson quasi-DA. A comprehensive theoretical analysis of both the IM and LMs in LaMET will be conducted, with a specific focus on presenting the one-loop mixing matrix. Additionally, a numerical analysis of the LMs of the quasi-DA will be performed.

The rest of this paper is organized as follows. We define the IM and LMs of $B$-meson quasi-DA in Sec.\,\ref{Definitions}. In Sec.\,\ref{Calculation}, we perform the one-loop calculation on IM and LMs of quasi-DA and LCDA, respectively; The discussion on factorization formula and a brief phenomenological analysis will  be given in Sec.\,\ref{FactorizationFormula}. The last section includes a brief summary and an outlook for future works.

\section{Inverse and logarithmic moments of quasidistribution amplitude}
\label{Definitions}
We follow the notations in \cite{Xu:2022krn}. The $B$-meson LCDA $\phi_B^+(\omega,\mu)$  in momentum space can be deduced from the Fourier transform of the LCDA in coordinate space
\begin{eqnarray}\label{Fourier1}
  \phi_B^+(\omega,\mu) = \frac{v^+}{2\pi} \int_{-\infty}^{+\infty} d\eta \, e^{i\omega v^+ \eta} \, \tilde\phi_B^+(\eta,\mu)  \,,
\end{eqnarray}
here $v_\mu$ is the heavy quark velocity satisfying $v^2=1$ and $v^+ \!\equiv n_+ \!\cdot\! v \,(v^- \!\equiv n_- \!\cdot\! v)$, with $n_{\pm}$ being the unit light-cone vectors
\begin{eqnarray}
	n_{+\mu}=\frac{1}{\sqrt{2}}(1,0,0,1) \,, \quad n_{-\mu}=\frac{1}{\sqrt{2}}(1,0,0,-1) \,.
\end{eqnarray}
The LCDA in coordinate space  $\tilde\phi_B^+(\eta,\mu)$  is defined through the renormalized HQET matrix element of a light-cone operator \cite{Lange:2003ff}
\begin{eqnarray}
  &&\left\langle 0\left|\bar{q}(\eta n_+) \slash{n}_+\gamma_5 W(\eta n_+, 0) h_v(0)\right| \bar{B}(v)\right\rangle \nn\\
  &&= i \tilde f_B(\mu) M \tilde\phi_B^+(\eta,\mu) v^+ \, ,
\end{eqnarray}
where $W(\eta n_+, 0)=\mathrm{P}\left\{\operatorname{Exp}\left[i g_{s} \int_{0}^{\eta} d x \, n_+ \!\cdot\! A(x n_+)\right]\right\}$ is Wilson line connecting the light and heavy quark fields, ensuring the gauge invariance;  $\tilde f_B(\mu)$ is the $B$-meson decay constant in HQET \cite{Beneke:2005gs}.
Therefore, $\phi_B^+(\omega,\mu)$ can be expressed in terms of the nonlocal and local matrix elements as
\begin{eqnarray}\label{defiofphi+}
  &&\phi_B^+(\omega,\mu) = v^+ \int_{-\infty}^{+\infty} \frac{d\eta}{2\pi} \, e^{i\omega v^+ \eta} \nn\\
   &&\quad \times  \frac{\left\langle 0\left|\bar{q}(\eta n_+) \slash{n}_+\gamma_5 W(\eta n_+, 0) h_v(0)\right| \bar{B}(v)\right\rangle}{\left\langle 0\left|\bar{q}(0) \slash{n}_+ \gamma_5 h_v(0)\right| \bar{B}(v)\right\rangle} \,.
\end{eqnarray}
The first IM of $B$-meson LCDA is defined as
\begin{eqnarray}\label{eqIMLCDA}
  \lambda_B^{-1}(\mu)\equiv \int_{0}^{\infty}d\omega \frac{\phi_{B}^+(\omega,\mu)}{\omega} \, ,
\end{eqnarray}
while the logarithmic moments are \cite{Braun:2003wx}
\begin{eqnarray}\label{eqlog}
  \sigma_n(\mu) \equiv \lambda_B(\mu)\int_{0}^{\infty}\frac{d\omega}{\omega}\ln^n \frac{\mu}{\omega} \phi_B^+(\omega,\mu) \,.
\end{eqnarray}

According to~\cite{Wang:2019msf}, the quasi-DA is defined through the equal-time matrix element in HQET,
\begin{eqnarray}\label{defiofquasiphi-}
  &&\varphi_B^+(\xi, v^z, \mu) = v^z \int_{-\infty}^{+\infty} \frac{d\tau}{2\pi} \, e^{i\xi v^z \tau} \nn\\
  &&\quad \times \frac{\left\langle 0\left|\bar{q}(\tau n_z) \slash{n}_z \gamma_5 W(\tau n_z, 0) h_v(0)\right| \bar{B}(v)\right\rangle}{\left\langle 0\left|\bar{q}(0) \slash{n}_z\gamma_5 h_v(0)\right| \bar{B}(v)\right\rangle} \,.
\end{eqnarray}
Here, $n_{z\mu}=(0,0,0,1)$ and $v^z \equiv n_z \!\cdot\! v$. We will work in a Lorentz boosted frame of the $B$-meson in which $v^+ \gg v^-$ and $v_{\perp\mu}=0$.  Because there is no time-dependence in Eq.~\eqref{defiofquasiphi-}, the quasi-DA can be simulated directly on the lattice.
Note that the support of quasi-DA is $(-\infty,\infty)$ while for LCDA the support is $[0,\infty)$.  As in~\cite{Xu:2022krn}, we define the first IM of quasi-DA as
\begin{eqnarray}\label{quasiIM}
  \widetilde{\lambda}_B^{-1}(v^z,\mu) \equiv \operatorname{P.V.}\int_{-\infty}^{\infty}d\xi \frac{\varphi_{B}^+(\xi, v^z, \mu)}{\xi}\, ,
\end{eqnarray}
and similarly, the logarithmic moments of quasi-DA can be defined as
\begin{align}\label{quasilog}
  \widetilde{\sigma}_n(v^z, \mu) & \equiv \widetilde\lambda_B(v^z, \mu_Q)  \nonumber\\
  &\!\!\!\!\!\!\times \operatorname{P.V.}\int_{-\infty}^{\infty}\frac{d\xi}{\xi}\ln^n \frac{\mu}{|\xi|} \varphi_B^+(\xi, v^z, \mu) \, ,
\end{align}
where the Cauchy principal value $\operatorname{P.V.}$ is introduced as a prescription of the singularities at $\xi=0$ in the integrands in Eqs.\,\eqref{quasiIM} and \eqref{quasilog}. One can also utilize other prescriptions, see, e.g., Ref.~\cite{Xu:2022krn}.

%
%
\section{One-loop results}
\label{Calculation}
To calculate the radiative corrections of IM and LMs at one-loop, we replace the $B$-meson state with a heavy $b$ quark plus an off-shell light quark. The off-shellness of the initial light quark serves as an infrared (IR) regulator, $k^2 = 2k^+ k^- - k_\perp^2= k^{t2} -k^{z2} - k_\perp^2$. The dimensional regularization ($d\!=\!4\!-\!2\epsilon$) with modified minimum subtraction scheme ($\overline{\textrm{MS}}$ scheme) is utilized in our calculation.
\begin{figure*}
  \centering
  \includegraphics[width=0.4\linewidth]{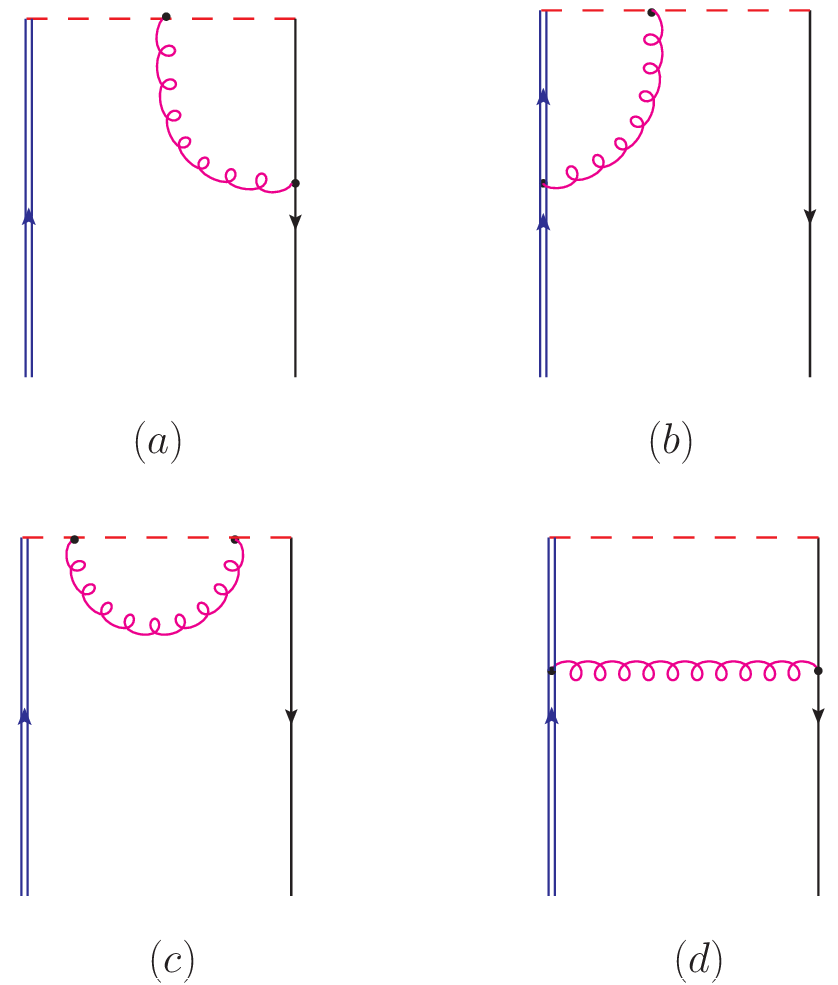}
  \caption{(Color online) The Feynman diagrams for calculating IM and LMs of quasi-DA and LCDA. The red dashed line represents the gauge link, while the blue double line denotes the heavy quark in HQET. The single line represents the light quark.}
  \label{diagram}
\end{figure*}

The calculation is performed in Feynman gauge.
The relevant  Feynman diagrams at one-loop are shown in Fig.\,\ref{diagram}. The results of the IM and the first two LMs of LCDA are:
\begin{widetext}
\begin{subequations}\label{LCDAMs}
\begin{eqnarray}
  \label{LCDAIM1}
  \lambda_B(\mu) &=& k^0 + \frac{\alpha_s C_F}{4\pi}k^0 \left[ 2\ln^2\frac{\mu}{k^0} -2\ln\frac{\mu^2}{-k^2} -4+\frac{3\pi^2}{4} \right]+\mathcal{O}(\alpha_s^2) \,,\\\nn\\
  \label{LCDAsigma1}
  \sigma_1(\mu)
  &=& \ln\frac{\mu}{k^0} +\frac{1}{k^0}\ln\frac{\mu}{k^0}\lambda_B^{(1)} \nonumber\\
  &&+\frac{\alpha_s C_F}{4\pi} \frac{1}{k^0} \bigg[-2\ln^3\frac{\mu}{k^0} +\frac{1}{12}\ln\frac{\mu}{k^0}\Big( 48-17\pi^2\Big)  +\left(2\ln\frac{\mu}{k^0}+\frac{\pi^2}{3}
  \right)\ln\frac{\mu^2}{-k^2}+10\zeta(3)\bigg] +\mathcal{O}(\alpha_s^2) \,,\\\nn\\
  \label{LCDAsigma2}
  \sigma_2(\mu)
  &=& \ln^2\frac{\mu}{k^0} +\frac{1}{k^0}\ln^2\frac{\mu}{k^0}\lambda_B^{(1)} -\frac{\alpha_s C_F}{4\pi}\frac{1}{k^0}\bigg[ 2\ln^4\frac{\mu}{k^0}+\ln^2\frac{\mu}{k^0}\Big( -4+\frac{25\pi^2}{12}\Big)  \nn\\
  && -\ln\frac{\mu^2}{-k^2}\left(2\ln^2\frac{\mu}{k^0}+\frac{2\pi^2}{3}\ln\frac{\mu}{k^0}
  +4\zeta(3)\right) -28\ln\frac{\mu}{k^0}\zeta(3)+\frac{\pi^4}{10} \bigg] +\mathcal{O}(\alpha_s^2) \,.
\end{eqnarray}
\end{subequations}
\end{widetext}
These moments are expressed in series of $\mathcal{O}(\alpha_s)$, here $k^0 = k^+/v^+ = k^z/v^z$, and $\zeta(s)$ is Riemann zeta function. The $k^2$ in logarithm serves as IR regulator which will cancel with the IR divergence in corresponding quasimoments.
Similarly, the results of IM and the first two LMs of quasi-DA are:
\begin{widetext}
\begin{subequations}\label{quasiMs}
\begin{eqnarray}
  \label{quasiIM1}
  \widetilde\lambda_B(v^z, \mu)
  &=& k^0 + \frac{\alpha_s C_F}{4\pi}k^0 \bigg[ 2\left(2\ln2v^z+1\right)\ln\frac{\mu}{k^0}-2\ln\frac{k^{02}}{-k^2} -2\ln2v^z\left(\ln 2v^z+3\right)+2+ \frac{5\pi^2}{6}\bigg]+\mathcal{O}(\alpha_s^2) \,,\\\nn\\
  \label{quasisigma1} \widetilde{\sigma}_1(v^z, \mu)
  &=& \ln\frac{\mu}{k^0} + \frac{1}{k^0}\ln\frac{\mu}{k^0} \, \widetilde\lambda_B^{(1)} \nn\\
  && +\frac{\alpha_s C_F}{4\pi} \frac{1}{k^0} \bigg[-2\left(2\ln2v^z+1\right)\ln^2\frac{\mu}{k^0} -\ln\frac{\mu}{k^0}\Big(-2\ln^22v^z-6\ln2v^z+ 2+\frac{11\pi^2}{6}\Big) \nn\\
  && +\Big(\frac{\pi^2}{3}+2\ln\frac{\mu}{k^0}\Big)\ln\frac{k^{02}}{-k^2} +\pi^2\left(\frac{1}{6}+\ln2v^z\right)+6\zeta(3) \bigg]+\mathcal{O}(\alpha_s^2) \,,\\\nn\\
  \label{quasisigma2}
  \widetilde{\sigma}_2(v^z, \mu)
  &=& \ln^2\frac{\mu}{k^0} + \frac{1}{k^0}\ln^2\frac{\mu}{k^0} \, \widetilde\lambda_B^{(1)} \nn\\
  &&-\frac{\alpha_s C_F}{4\pi} \frac{1}{k^0} \bigg[2\left(2\ln2v^z+1\right)\ln^3\frac{\mu}{k^0} +\left(-2\ln^22v^z-6\ln2v^z +2+\frac{17\pi^2}{6}\right)\ln^2\frac{\mu}{k^0}\nonumber\\
  &&-\left(2\pi^2\ln2v^z+\frac{\pi^2}{3}\right)\ln\frac{\mu}{k^0} -4\zeta(3)\left(\ln\frac{\mu^2}{-k^2}+\ln\frac{\mu}{k^0}\right)-\ln\frac{\mu}{k^0}\ln\frac{(k^{0})^2}{-k^2}\left(
  2\ln\frac{\mu}{k^0}+\frac{2\pi^2}{3}\right)\nonumber\\
  &&-8\zeta(3)\left(2\ln2v^z+1\right)+\frac{7\pi^4}{30}\bigg]+\mathcal{O}(\alpha_s^2)  \,.
\end{eqnarray}
\end{subequations}
\end{widetext}
It is worth noting that these moments of quasi-DA changes dynamically under a boost along $z$ direction, whose dependence is encoded in the nontrivial expressions of the heavy quark velocity $v$.

\section{Factorization formula in LaMET}
\label{FactorizationFormula}
In LaMET, the quasi-quantities can be linked to their light-cone counterparts via a matching formula. Given that neither the moments of LCDA nor quasi-DA involve dependencies on $\omega$ and $\xi$, the matching relations would in the form of multiplicative relationships instead of convolutions. In \cite{Xu:2022krn},  the factorization formula was written down for IM of quasi-DA by observing that the  $\ln^n(\mu/k^0)$ terms in $\lambda_B(\mu)$ and $\widetilde\lambda_B(v^z, \mu)$ are related to the logarithmic moments defined in Eq.\,(\ref{eqlog}),
\begin{eqnarray}\label{facformula1}
  \widetilde{\lambda}_B (v^z, \mu) &=& \lambda_B(\mu) \Big[ C_0 \left(v^z\right) +C_1 \left(v^z\right) \sigma_1\left(\mu\right) \nn\\
  && +C_2 \left(v^z\right) \sigma_2\left(\mu\right) \Big] +\mathcal{O}(1/v^z) \,.
\end{eqnarray}
Performing the perturbative expansion, at tree level, we have $C_0^{(0)} = 1 \,, C_{i\neq0}^{(0)}=0.$ At one-loop, making use of Eqs.\,(\ref{LCDAIM1}) and (\ref{quasiIM1}), one has
\begin{eqnarray}\label{coeset1}
  C_0^{(1)}\left(v^z\right) &=& \frac{\alpha_s C_F}{4\pi}\Big( -2\ln^2 2v^z -6\ln2v^z +\frac{\pi^2}{12} +6 \Big) \,,\nn\\
  C_1^{(1)}\left(v^z\right) &=& \frac{\alpha_s C_F}{4\pi}\Big( 4\ln2v^z+6 \Big) \,,\nn\\
  C_2^{(1)}\left(v^z\right) &=& \frac{\alpha_s C_F}{4\pi}\Big(-2 \Big) \,.
\end{eqnarray}
This result is consistent with the one in \cite{Xu:2022krn}. For the quasilogarithmic moment $\widetilde\sigma_1$, the  matching formula is expected to be
\begin{align}\label{facformula2}
 & \widetilde{\sigma}_1 (v^z, \mu) = \frac{\lambda_B(\mu)}{\widetilde\lambda_B(v^z, \mu)} \Big[ B_0\left(v^z\right) +B_1\left(v^z\right)\sigma_1\left(\mu\right) \nn\\
  &+B_2\left(v^z\right)\sigma_2\left(\mu\right) +B_3\left(v^z\right)\sigma_3\left(\mu\right) \Big] +\mathcal{O}(1/v^z) \,.
\end{align}
The tree level result of matching coefficients $B_i$s are $B_1^{(0)}=1 \,, B_{i\neq1}^{(0)} =0.$ According to Eqs.\,(\ref{LCDAsigma1}) and (\ref{quasisigma1}), the one-loop result reads
\begin{eqnarray}\label{coeset2}
  B_0^{(1)}\left(v^z\right) &=& \frac{\alpha_s C_F}{4\pi}\Big( \pi^2 \ln2v^z -4\zeta(3) +\frac{\pi^2}{6} \Big)  \,, \nn\\
  B_1^{(1)}\left(v^z\right) &=& \frac{\alpha_s C_F}{4\pi}\Big( -2\ln^2 2v^z -6\ln2v^z -\frac{11}{12}\pi^2 +6 \Big)  \,, \nn\\
  B_2^{(1)}\left(v^z\right) &=& \frac{\alpha_s C_F}{4\pi}\Big( 4\ln 2v^z +6 \Big)  \,, \nn\\
  B_3^{(1)}\left(v^z\right) &=& \frac{\alpha_s C_F}{4\pi}\Big(-2 \Big)  \,.
\end{eqnarray}
Similarly, the matching formula for quasilogarithmic moment $\widetilde\sigma_2$ is
\begin{eqnarray}\label{facformula3}
  \widetilde{\sigma}_2(v^z, \mu) &=& \frac{\lambda_B(\mu)}{\widetilde\lambda_B(v^z, \mu)} \Big[ A_0\left(v^z\right) +A_1\left(v^z\right)\sigma_1\left(\mu\right) \nn\\
  && +A_2\left(v^z\right)\sigma_2\left(\mu\right)+A_3\left(v^z\right)\sigma_3\left(\mu\right) \nn\\
  && +A_4\left(v^z\right)\sigma_4\left(\mu\right) \Big] +\mathcal{O}(1/v^z) \,.
\end{eqnarray}
At tree level, we have $A_2^{(0)}=1 \,, A_{i\neq2}^{(0)} =0.$ The next-to-leading order corrections of these hard coefficients are
\begin{eqnarray}\label{coeset3}
  A_0^{(1)}\left(v^z\right) &=& \frac{\alpha_s C_F}{4\pi}\Big( 16\zeta(3)\ln 2v^z +8\zeta(3) -\frac{2}{15}\pi^4 \Big) \,, \nn\\
  A_1^{(1)}\left(v^z\right) &=& \frac{\alpha_s C_F}{4\pi}\Big( 2\pi^2\ln 2v^z -24\zeta(3) +\frac{\pi^2}{3} \Big) \,, \nn\\
  A_2^{(1)}\left(v^z\right) &=& \frac{\alpha_s C_F}{4\pi}\Big( -2\ln^2 2v^z -6\ln 2v^z -\frac{23}{12}\pi^2+6 \Big) \,, \nn\\
  A_3^{(1)}\left(v^z\right) &=& \frac{\alpha_s C_F}{4\pi}\Big( 4\ln 2v^z +6 \big) \,, \nn\\
  A_4^{(1)}\left(v^z\right) &=& \frac{\alpha_s C_F}{4\pi}\Big( -2 \Big) \,.
\end{eqnarray}
The effectiveness of LaMET relies on that the moments of quasi-DA and LCDA share the exactly same IR properties. It indicates that the matching coefficients do not depend on the IR regulator $\ln(-k^2)$, as well as $k^0$ which is related to the momentum of external light quark. One can see in Eqs.\,(\ref{coeset1}), (\ref{coeset2}) and (\ref{coeset3}) the matching coefficients fulfill these requirements.

Another intriguing observation is that $C_1^{(1)}$, $B_2^{(1)}$, and $A_3^{(1)}$ exhibit remarkable identicality at one-loop level. The same thing happens with $C_2^{(1)}$, $B_3^{(1)}$ and $A_4^{(1)}$. The reason behind it is that these coefficients represent mixing with higher-order power of logarithmic moments (the IM can be regarded as the zeroth logarithmic moment). We find that for a certain quasilogarithmic moment $\widetilde\sigma_n$, the mixing with higher-power logarithmic moments $\sigma_{n+1}$, $\sigma_{n+2}$ is determined by the $\lambda_B(\mu)/\widetilde\lambda_B(v^z, \mu)$ term in the factorization formulas Eqs.\,(\ref{facformula1}), (\ref{facformula2}) and (\ref{facformula3}). The appearance of higher-power logarithmic moments during the calculations on $\widetilde{\sigma}_n(v^z, \mu)$ and $\sigma_n(v^z, \mu)$ at one-loop will cancel each other out.

Based on these analyses, a complete mixing matrix can be written down
\begin{widetext}
\begin{eqnarray}\label{facformulaG}
\begin{pmatrix}
  1  \\
  \widetilde{\sigma}_1  \\
  \widetilde{\sigma}_2  \\
  \cdot  \\
  \cdot  \\
  \widetilde{\sigma}_n  \\
\end{pmatrix}
= \,\frac{\lambda_B}{\widetilde\lambda_B}
\begin{pmatrix}
  H_{1,1} & H_{1,2} & H_{1,3} & 0 & 0 & \cdot & \cdot & 0 & 0 & 0  \\
  H_{2,1} & H_{2,2} & H_{2,3} & H_{2,4} & 0 & \cdot & \cdot & 0 & 0 & 0  \\
  H_{3,1} & H_{3,2} & H_{3,3} & H_{3,4} & H_{3,5} & \cdot & \cdot & 0 & 0 & 0  \\
  \cdot & \cdot & \cdot & \cdot & \cdot & \cdot & \cdot & \cdot & \cdot & \cdot  \\
  \cdot & \cdot & \cdot & \cdot & \cdot & \cdot & \cdot & \cdot & \cdot & \cdot  \\
  H_{n,1} & H_{n,2} & H_{n,3} & H_{n,4} & H_{n,5} & \cdot & \cdot & H_{n,n} & H_{n,n+1} & H_{n,n+2}
\end{pmatrix}
\times
\begin{pmatrix}
  1  \\
  \sigma_1  \\
  \sigma_2  \\
  .  \\
  .  \\
  \sigma_n  \\
  \sigma_{n+1}  \\
  \sigma_{n+2}  \\
\end{pmatrix}  \,,
\end{eqnarray}
with Eqs.\,(\ref{facformula1}), (\ref{facformula2}) and (\ref{facformula3}) as its first three rows. Some of the matching coefficients have been calculated or deduced at one-loop,
\begin{eqnarray}
\left\{
  \begin{array}{ll}
    H_{1,1}=C_0^{(0)} + C_0^{(1)} &   \\
    H_{1,2}=C_1^{(1)} &   \\
    H_{1,3}=C_2^{(1)} &
  \end{array}
\right. ,\,\,
\left\{
  \begin{array}{ll}
    H_{2,1}=B_0^{(1)} &   \\
    H_{2,2}=B_1^{(0)}+B_1^{(1)} &   \\
    H_{2,3}=B_2^{(1)} &   \\
    H_{2,4}=B_3^{(1)} &
  \end{array}
\right. ,\,\,
\left\{
  \begin{array}{ll}
    H_{3,1}=A_0^{(1)} &   \\
    H_{3,2}=A_1^{(1)} &   \\
    H_{3,3}=A_2^{(0)}+A_2^{(1)} &   \\
    H_{3,4}=A_3^{(1)} &   \\
    H_{3,5}=A_4^{(1)} &
  \end{array}
\right. ,\,\, \rm{and}~
\left\{
  \begin{array}{ll}
    H_{i,i+1}= \frac{\alpha_s C_F}{4\pi}( 4\ln 2v^z +6 ) &   \\\\
    H_{i,i+2}= \frac{\alpha_s C_F}{4\pi}( -2 ) &
  \end{array}
\right. .\nn\\
\end{eqnarray}
\end{widetext}

The matching formula in Eq.\,(\ref{facformulaG}) may provide a convenient approach to extract the inverse and logarithmic moments of $B$-meson LCDA. $\widetilde\lambda_B$ and $\widetilde{\sigma}_n$ can be expressed as  polynomials of $\ln 2v^z$, with coefficients involving the light-cone $\lambda_B$ and $\sigma_n$ according to Eq.\,(\ref{facformulaG}). Therefore the light-cone moments can be extracted by a polynomial fit of quasimoments with several different values of $v^z$.

\begin{figure}[h]
	\centering
	\includegraphics[width=0.75\columnwidth]{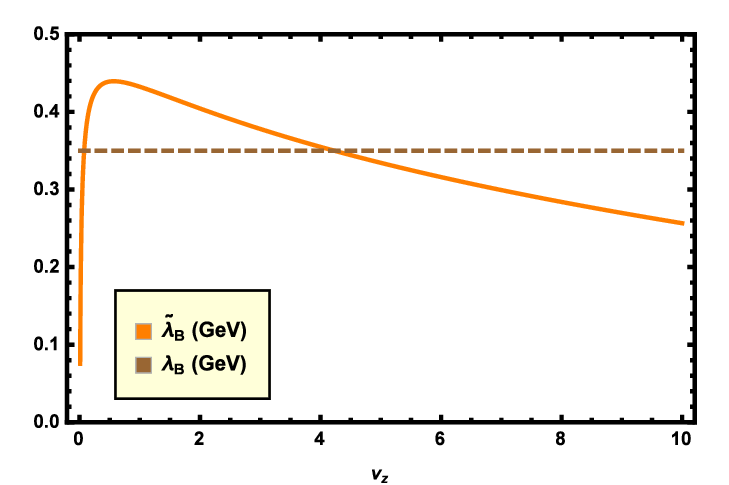}
	\includegraphics[width=0.75\columnwidth]{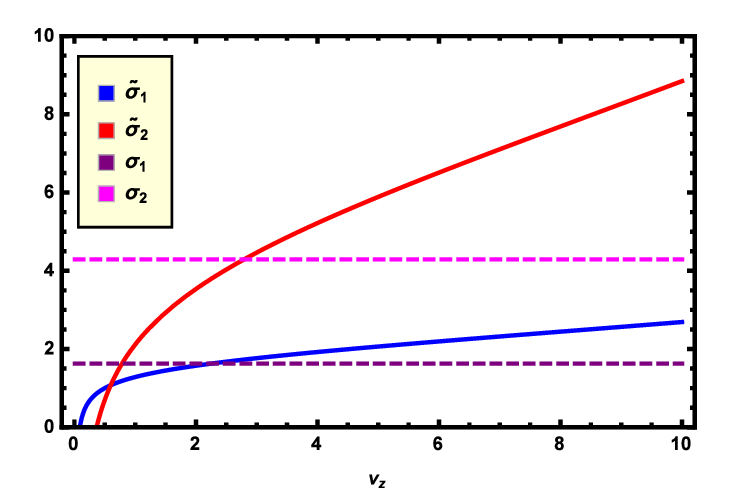}
	\centering
	\caption{ The inverse moment (upper panel) and logarithmic moments (lower panel) of quasi-DA as functions of $v^z$, obtained from numerical values of inverse and logarithmic moments of LCDA through HQET sum rule and factorization formula in Eq.\,(\ref{facformulaG}).}
	\label{diagram12}
\end{figure}

Due to the current absence of nonperturbative simulations for quasimoments $\widetilde\lambda_B$ and $\widetilde{\sigma}_n$, it is valuable to examine their characteristics through the lens of phenomenological models.
Starting with values of light-cone moments $\lambda_B$ and $\sigma_n$ calculated by the HQET sum rules in Table~\ref{tabmoments}, we want to find out to which extent the quasimoments will resemble their corresponding light-cone ones in magnitude.
\begin{table}[!htbp]
	\caption{Numerical values of inverse and logarithmic moments in terms of HQET sum rule \cite{Grozin:1996pq,Lee:2005gza,Braun:2003wx}. The scale is fixed to be $\mu=1\,\rm{GeV}$.}\label{tabmoments}
	\scalebox{1.15}{\begin{tabular}{c c c c c}
		\hline\hline
		$\,\quad\lambda_B\quad\,$ & $\,\quad\sigma_1\quad\,$ & $\,\quad\sigma_2\quad\,$ & $\,\quad\sigma_3\quad\,$  & $\,\quad\sigma_4\quad\,$ \\\hline
		$0.35 \,\rm{GeV}$ & 1.63 & 4.30 & 14.74 & 63.39  \\
		\hline\hline
	\end{tabular}}
\end{table}
The factorization formula in Eq.\,(\ref{facformulaG}) implies the values of quasimoments. The magnitudes of $\widetilde{\lambda}_B$, $\widetilde{\sigma}_1$ and $\widetilde{\sigma}_2$ as a function of $v^z$ are presented in Fig.~\ref{diagram12}, with $\mu$ fixed at 1\,GeV.
We note that the matching formula in Eq.\,(\ref{facformulaG})  only holds when $v^z$ is sufficiently large. On the other hand, when $v^z$ is too large, the large double and single logarithms of $v^z$ may weaken the convergence of perturbative expansion,  hence a resummation is required.

\section{Conclusion}\label{Conclusion}
In this work, we generalize a previous work~\cite{Xu:2022krn} on inverse momentum of $B$-meson LCDA to the logarithmic moments, which are among the most phenomenologically significant nonperturbative quantities in $B$-meson physics.
We introduce the inverse and logarithmic moments of quasi-DA and explore their properties. They can be simulated on a Euclidean lattice and can be expressed as the linear combinations of light-cone moments, in which the coefficients are calculable in perturbation theory. The mixing pattern in the matching formula in Eq.\,(\ref{facformulaG}) is presented and the matching coefficients are determined at one-loop.
%
The findings presented in this study represent a incremental stride toward a more comprehensive understanding of the matching properties within the framework of LaMET. These results pave the way for future realistic lattice studies focusing on the inverse moment and logarithmic moments of the $B$-meson LCDA.

\section*{Acknowledgements}
The authors thank Wei Wang for fruitful discussions. S. M. H and J. X. are supported by National Natural Science Foundation of China under Grant No. 12105247, the China Postdoctoral Science Foundation under Grant No. 2021M702957.

\end{document}